\documentclass[]{article} 

\usepackage{arxiv}
\usepackage{graphicx}

\newcommand{\radu}{rad m$^{-2}$}

\newcommand{\inv}{$^{-1}$}

\newcommand*\arcmin{\ensuremath{^\prime}}
\newcommand*\arcsec{\ensuremath{^{\prime\prime}}}
\usepackage{upgreek}

\def\arxivprefixesep{:}
\newcommand{\eprint}[2][]{%
{\tt\if!#1!#2\else#1\arxivprefixesep\ignorespaces#2\fi}%
}

\pdfoutput=1

\title{The Power of Low Frequencies: Faraday Tomography in the sub-GHz regime}

\newcommand\shorttitle{Faraday Tomography in the sub-GHz regime}
\newcommand\shortauthor{C.L.~Van Eck}

\author{Cameron L. Van Eck\\
Dunlap Institute for Astronomy and Astrophysics, \\University of Toronto, 50 St. George Street, Toronto, ON M5S 3H4, Canada \\ Correspondence email: cameron.van.eck@dunlap.utoronto.ca\\
Department of Physics and Astronomy, University of Calgary, Calgary, Alberta, T2N 1N4, Canada}

\date{}

\begin{document}

\maketitle

\begin{abstract}
Faraday tomography, the study of the distribution of extended polarized emission by strength of Faraday rotation, is a powerful tool for studying magnetic fields in the interstellar medium of our Galaxy and nearby galaxies. The strong frequency dependence of Faraday rotation results in very different observational strengths and limitations for different frequency regimes. I discuss the role these effects take in Faraday tomography below 1 GHz, emphasizing the 100-200 MHz band observed by the Low Frequency Array and the Murchison Widefield Array. With that theoretical context, I review recent Faraday tomography results in this frequency regime, and discuss expectations for future observations.
\end{abstract}


\section{Introduction}
The study of magnetic fields in astrophysical systems has been a topic of great interest for many years as, to paraphrase \citet{Woltjer1967}, it has often been the case that aspects of observations that cannot be explained by models are blamed on the effects of the (unmodelled) magnetic field. While many of the basic methods of observing astrophysical magnetic fields have been used for several decades, such as synchrotron polarization \citep{Gardner1962}, Faraday rotation \citep{Cooper1962}, starlight polarization \citep{Hiltner1949}, polarized dust emission \citep{Stein1966}, and the Zeeman effect \citep{Hale1908}, recent technological advances have drastically increased the amount of information at our disposal.

Faraday rotation measurements in particular have become considerably more flexible and powerful with the development of rotation measure (RM) synthesis \citep{Burn1966,Brentjens05}. RM synthesis allows for measured polarized signals to be separated by their strength of Faraday rotation, called the Faraday depth, but requires large bandwidths with many frequency channels in order to be effective and is still subject to a number of limitations. The power of RM synthesis is now accessible through the combination of broad-band receivers and high-capacity correlators/back-ends that have been developed in recent years for new radio telescopes, such as the Low Frequency Array (LOFAR) \citep{vanHaarlem2013} or the Australian Square-Kilometer Array Pathfinder (ASKAP) \citep{Johnson2007} and as upgrades to existing telescopes, such as the Karl G. Jansky Very Large Array (VLA). However, with the use of RM synthesis to analyze radio polarization measurements also comes the need to understand the properties and limitations of the RM synthesis method, as these affect how we can interpret the results and what information we can extract on the underlying physics, particularly the magnetic field in the region being studied. A detailed review the physical motivations for studying radio polarization and Faraday rotation, particularly in the Milky Way and other spiral galaxies, can be found in \citet{Beck2015}.

The Faraday depth, $\phi(x)$, describes the strength of Faraday rotation experienced by polarized emission at a distance $x$ from the observer. It is defined as 
\begin{equation}
\phi(x) =  0.812\; {\rm rad \, m^{-2}} \int_x^0 \left( \frac{n_\mathrm{e}(l)}{{\rm cm^{-3}}}\right) \left( \frac{\vec{B}(l)}{{\rm \upmu G}} \right) \cdot \left(\frac{\vec{dl}}{{\rm pc}} \right)
\end{equation}
where $n_e(l)$ is the local thermal electron density, $\vec{B}(l)$ is the magnetic field, and the product of these is integrated over the position $l$ along the line of sight, from $x$ to the observer. The Faraday depth, multiplied by the observing wavelength, $\lambda$, squared gives the rotation of the polarization angle for polarized emission at that wavelength and distance. For lines of sight with only a single (dominant) source of polarized emission at some distance, the measured polarization angle follows a linear relationship with $\lambda^2$; prior to RM synthesis the most common method of measuring Faraday depth was to determine the slope of this linear relationship \citep[e.g.,][]{Rand1994,Brown2003}, which was called the rotation measure and equal to the Faraday depth only in this simple case. More generally, the presence of multiple sources of polarized emission at different distances, or polarized emission that is distributed over distance with internal Faraday rotation along that distance, for a given line of sight will result in the superposition of all these polarized signals being observed. This superposition will generally not retain this linear relationship, causing the RM to not accurately reflect the Faraday depth of any of the polarized emission.

RM synthesis provides more flexibility, as it exploits the mathematical properties of polarization and Faraday rotation to make a Fourier-like transformation between the observed polarization as a function of wavelength squared to the polarization as a function of Faraday depth. In this way, multiple sources of emission with different Faraday depths can potentially be distinguished from each other (subject to limitations in the effectiveness of the transformation). The output of RM synthesis, for a given line of sight, is a Faraday depth spectrum, giving the polarization for every possible Faraday depth value. When this is applied to an entire field-of-view, for the study of extended or diffuse polarized emission, it is often called Faraday tomography: the use of Faraday depth cubes (two dimensions spanning the image plane, with the third spanning Faraday depth) to analyze Faraday rotation along potentially complex lines of sight. As a Fourier transform-based technique, RM synthesis (and by extension Faraday tomography) has a number of properties and limitations, including a point spread function and corresponding limited resolution in Faraday depth, that are very similar to those found in radio interferometry; these are a major subject of consideration in this article and will be described in detail below. The strong frequency dependence of Faraday rotation affects many of these properties, so even relatively moderate changes in frequency can have very significant effects on the use of RM synthesis, and in turn on the measurable physical properties of the astrophysical systems being studied.

While early use of Faraday tomography was focused on interferometer observations at frequencies below 1 GHz, such as \citet{deBruyn2005} and \citet{Brentjens2011} who used data from the Westerbork Synthesis Radio Telescope (WRST) around 350 MHz, a lot of later work has been done at higher frequencies, such as \citet{Heald2009, Harvey-Smith2010, Mao2015} around 1.4 GHz; \citet{Mulcahy2017,Basu2017} at 2--4 GHz; and \citet{Irwin2017} around 6 GHz. \citet{Mao2012b} combined single-dish and interferometer data at 1.4 GHz, and \citet{Sun2015} and \citet{Hill2017} used single dish data at 1.4 GHz, to explore larger regions of the sky.
However, these different datasets must be interpreted very differently, as the different frequency coverage drastically changes the volume of interstellar medium being probed, as will be described in the next section.

In this proceedings, I review the current state of Faraday tomography at low frequencies, which I define here as below 1 GHz, with a focus on the 100-200 MHz frequency range observed by LOFAR and the Murchison Widefield Array \citep[MWA,][]{Tingay2013}. I begin by examining the key parameters affecting the limitations of RM synthesis, emphasizing their behaviour at low frequency and describing in detail how this affects the physical interpretation of the observed RM spectra, in Section~\ref{sec:theory}. With that context, I review the published work using low-frequency Faraday tomography in Section~\ref{sec:current} and discuss prospective future observations in Section~\ref{sec:future}. Section~\ref{sec:summary} provides a summary of the article.

\section{Theoretical Considerations}\label{sec:theory}

Since RM synthesis is based on the Fourier transform, most of its properties follow directly, such as shape of the point spread function, aliasing, convolution, and transform pairs. Many of these properties were laid out in a general form in \citet{Brentjens05}, so below I focus on just a few of these properties that show strongly frequency-dependent behaviour that may have differences between the low-frequency (sub-GHz) regime and higher frequencies.

\subsection{Properties of RM synthesis at low frequencies}
There are three key frequency-dependent parameters that determine the limitations of RM synthesis: the Faraday depth resolution, the largest scale in Faraday depth that can be measured, and the largest Faraday depth that can be measured. For observations with a single continuous block of bandwidth, these parameters are given by equations 61-63 of \citet{Brentjens05}. 

The Faraday depth resolution, defined as the width of the Rotation Measure Spread Function (RMSF), is inversely proportional to the range of $\lambda^2$ values measured. As a result, low frequency observations with even modest bandwidth will usually have much better Faraday depth resolution than broadband higher frequency observations; an observation from 1--2 GHz, possible with the VLA L-band receivers has a Faraday depth resolution of 50 \radu, while a typical observation in the LOFAR high-band from 110--170 MHz has a resolution of 1 \radu, under the assumption no bandwidth is lost to radio frequency interference. While it is possible to identify Faraday depth structure on scales smaller than the resolution \citep{Brown2017}, it is difficult to measure this structure unambiguously \citep{Kumazaki2014} and it requires high signal-to-noise. Low frequency observations are thus very desirable for studies of objects with small Faraday depth variations, such as giant radio galaxies or local interstellar medium (ISM) structure in the Milky Way.

The largest measurable scale in Faraday depth is a critical parameter for Faraday tomography, as it determines what types of polarization features will be observable. Many early polarization studies relied on the assumption that the polarized emission and Faraday rotation occurred in physically separated regions, so that all the emission experienced the same Faraday rotation and had a single, well-defined Faraday depth; this was called the Faraday-thin, one-component regime \citep{Vallee1980}. However, in many regions the polarized emission region is mixed with the Faraday rotation region, causing the polarized emission to be distributed over a range of Faraday depths. As a result, Faraday depth spectra can contain a mixture of very narrow (`Faraday-thin') components and broader components. Due to the Fourier transform relationship between the Faraday depth spectrum and the polarization as a function of wavelength (squared), these broad components have more intensity at shorter wavelengths and become depolarized at longer wavelengths (this can be called wavelength-dependent depolarization or depth depolarization, depending on whether the speaker is more oriented towards the wavelength or Faraday depth domain). 

Correspondingly, the largest measured scale in a Faraday depth spectrum is inversely proportional to the shortest $\lambda^2$ value observed, or equivalently, proportional to the highest frequency in an observation. More specifically, the largest scale is approximately equal to (adapting Eq. 62 of \citet{Brentjens05}) 35 \radu\ $(\nu_{\mathrm{max}}/\mathrm{GHz})^2$. Thus, a VLA L-band observation will be sensitive to scales smaller than 140 \radu, while a LOFAR-high observation is only sensitive to scales of 1.0 \radu\ and smaller. The largest scales present in a given astrophysical environment are expected to be approximately the same as the largest Faraday depth contribution produced in that environment, so environments like radio galaxies or galactic disks may have Faraday depth features with widths of several hundred or even a few thousand \radu\ \citep{Anderson16}, while local features or regions of the Galactic halo may have features with scales of up to a few tens of \radu\ \citep{Mao2012a}. As a result, low-frequency observations (without supporting higher-frequency data) will be insensitive to much of the polarized emission that we expect from these environments. However we do expect, and observe, a population of Faraday-thin features; I will discuss this in Section~\ref{sec:thin}.

\begin{figure}[htbp]
   \centering
   \includegraphics[width=\linewidth]{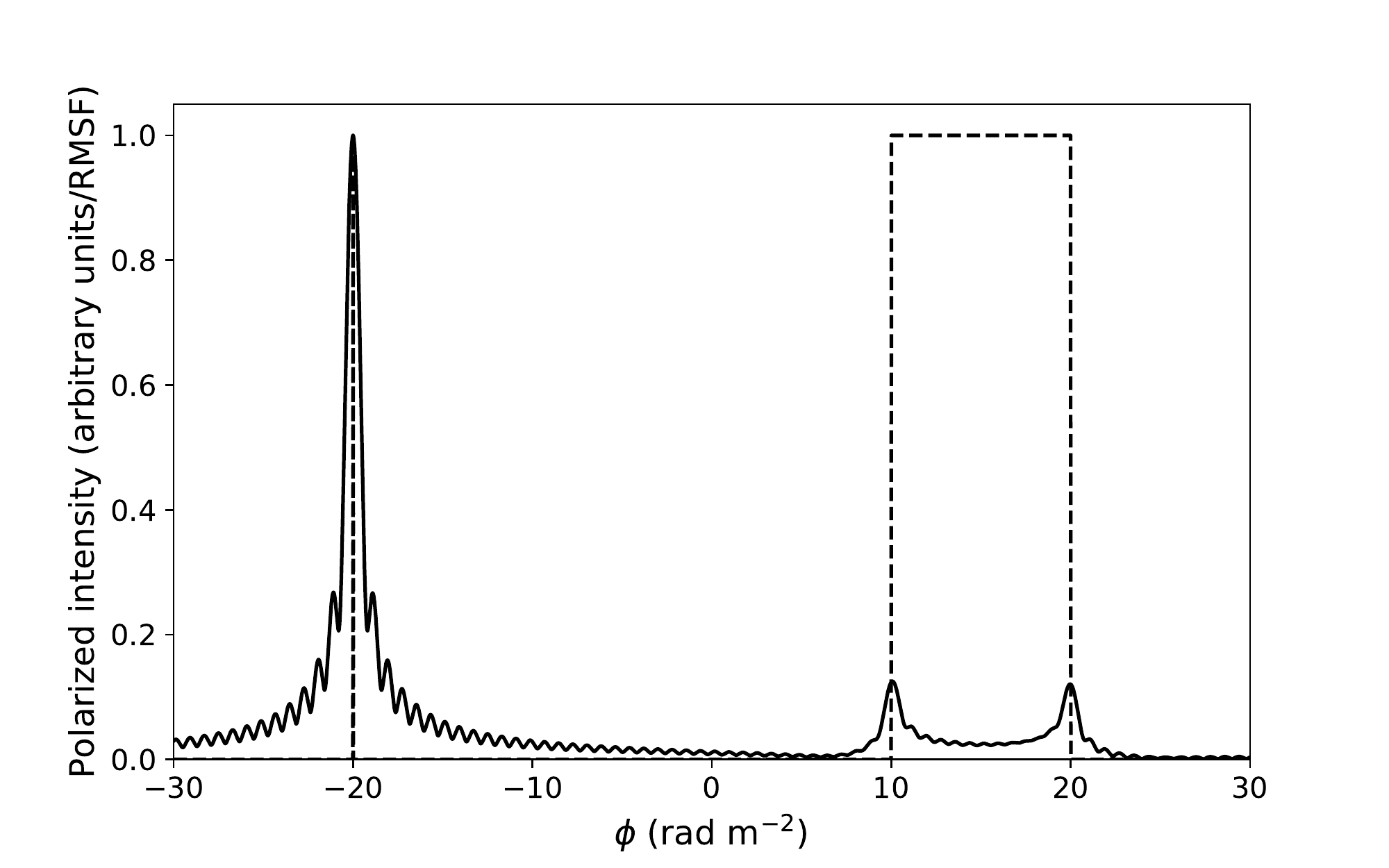} 
   \caption{A simulation of RM synthesis, where a simulated spectrum (dashed line) containing a Faraday-thin (delta-function) component and a Burn slab (top-hat) has been `observed' with the frequency coverage (115--178 MHz) and channel widths (24.4 kHz) of a sample LOFAR observation \citep{VanEck2017a}. In the resulting RM spectrum (solid line), the Faraday component takes on the shape of the RMSF, while the broad slab is mostly filtered out due to being significantly broader than the largest measured scale (1.1 \radu). Some emission is left at the edges of the slab, producing a characteristic `two-horn' shape.}
   \label{fig:simulator}
\end{figure}

Figure~\ref{fig:simulator} demonstrates the effects of the RMSF and the largest measured scale. The narrow feature in the left of the figure takes on the shape of the RMSF, while the broad feature on the right is much broader than the largest measured scale (1.1 \radu) is strongly depolarized except for the sharp edges, which are partially recovered. This is due to the Fourier property that sharp edges produce very broad features in the Fourier-conjugate domain ($\lambda^2$, in RM synthesis) which equates to some polarized emission remaining at long wavelengths; the result is that a Burn slab \citep{Burn1966} that is significantly broader than the largest scale will produce a `two-horned' Faraday depth feature \citep{Beck2012,VanEck2017a}.

It is worth noting that the largest measured scale can be, and often is, smaller than the Faraday depth resolution. As a result, it's possible to have observing configurations where no resolved features can be detected. Since the resolution depends (inversely) on the range of wavelength squared observed, $\Delta\lambda^2$, and the largest scale depends (inversely) on the shortest wavelength squared, $\lambda^2_\mathrm{min}$, we can express this in terms of the ratio of highest and lowest frequency. Equation 64 of \citet{Brentjens05} gives the condition to be able to resolve Faraday depth features as $\lambda^2_\textrm{min} < \Delta\lambda^2$; this can be re-expressed in terms of the ratio of highest and lowest frequencies as $\nu_\textrm{max}/\nu_\textrm{min} > \sqrt{2}$. For many modern radio telescopes, the large fractional bandwidth needed to meet this criterion is now available. Typical LOFAR-high observations barely satisfy this criterion, so only a very narrow range of scales could be resolved, and typical MWA observations \citep[e.g.,][]{Lenc2016} do not satisfy this criterion. Correspondingly, polarization features observed with current low-frequency instruments are almost always unresolved in Faraday depth, unless combined with additional observations to extend the frequency coverage.

The third limiting parameter of RM synthesis is the largest Faraday depth that can be measured. This limitation is caused by bandwidth depolarization, as very large (positive or negative) Faraday depths will cause significant change in polarization across the bandwidth of the channel. Therefore, the Faraday depth at which the sensitivity has dropped to 50\% is inversely proportional to the channel width in $\lambda^2$. For low-frequency observations, where small changes in frequency will produce larger changes in wavelength, it becomes important to ensure that channel widths are kept as small as possible to maximize sensitivity to larger RMs. This is reasonably straightforward for dedicated polarimetric observations (but can introduce data volume challenges, as the number of channels becomes large), but can be a problem with observations that were not originally optimized for polarization analysis \citep[e.g.][]{VanEck2018a}. Typical values for this parameter are harder to define, as most instruments allow the observer to select the channel width to match their own preferences; observations for the LOFAR Two-meter Sky Survey \citep[LOTSS, ][]{Shimwell2017} have a largest Faraday depth of 170 \radu\ (at the bottom of the band; the higher frequencies within the observing bandwidth will have lower depolarization and correspondingly higher largest Faraday depths) due to channel-averaging, observations from the Galactic and Extragalactic All-sky MWA survey (\citep[GLEAM, ][]{Wayth2015}) have a largest Faraday depth of 1937 \radu\ \citep{Riseley2018}, and the fine-spectral resolution data from the VLA Sky Survey will have a largest Faraday depth of 4800 \radu\ (at the bottom of the band). It is possible to correct for the reduced sensitivity at large Faraday depths \citep{Schnitzeler2015a}, at the cost of increased noise in those parts of the Faraday depth spectrum.

\subsection{Producing Faraday-thin features}\label{sec:thin}

As low-frequency Faraday tomography is primarily sensitive to Faraday-thin polarization features, the interpretation of such data requires an understanding of the physical conditions that can lead to such features being produced, as well as the conditions that will produce features too broad to be observed.
The criteria for defining such features were discussed by \citet{VanEck2017a}; a slightly more careful derivation of that result appears here.

The physical mapping between the polarized emission as a function of distance, $\tilde{P}(x)$,\footnote{All quantities with a tilde denote complex quantities, specifically polarization phasors.} and the Faraday depth spectrum, $\tilde{F}(\phi)$, can be written as \citep{Bell2011}:
\begin{equation}
\tilde{F}(\phi) = \int_0^\infty \delta(\phi-\phi(x)) \tilde{P}(x) dx,
\end{equation}
where $\phi(x)$ is the Faraday depth as a function of distance along the line of sight.\footnote{It is important to note that while both $\phi$ and $\phi(x)$ are called the Faraday depth, $\phi$ is the independent variable in the Faraday depth spectrum and spans the range of possible amounts of Faraday rotation ($-\infty$ to $\infty$ \radu), while $\phi(x)$ is the specific Faraday depth that would be observed for polarized emission produced at a given distance.}
The Dirac delta function in this equation can be interpreted as selecting all distances that produce the same Faraday depth, allowing them to be co-added to the same Faraday depth in the resulting Faraday depth spectrum; this is necessary to account for the possibility that multiple distances along a line of sight can have the same Faraday depth. Exploiting a property of the Dirac delta with a function as the argument, this can be rewritten as 
\begin{equation}\label{eq:amplitude}
\tilde{F}(\phi) = \sum_{x_\phi} \frac{\tilde{P}(x_\phi)}{\left| \frac{d}{d x}\phi(x)\right|_{x=x_\phi}} = \sum_{x_\phi} \frac{\tilde{P}(x_\phi)}{n_e(x_\phi) \left|B_\parallel(x_\phi)\right|}
\end{equation}
where $x_\phi$ is the set of distances that satisfy $\phi-\phi(x_\phi)=0$, and we have replaced the derivative of Faraday depth in the denominator with the free electron density and parallel magnetic field. The denominator in this equation represents the physical effect where stronger Faraday rotation causes the polarized emission to be more spread out in Faraday depth and correspondingly lowers the amplitude at any given single Faraday depth.

A Faraday-thin feature implies a narrow peak in the Faraday depth spectrum, which in turn requires that the right-hand side of Eq.~\ref{eq:amplitude} be much larger in a small range of Faraday depths compared to the rest of the Faraday depth spectrum. However, due to the high-pass filter effects of low-frequency observations there are two additional scenarios in which an observed Faraday spectrum mimicking a Faraday-thin feature may occur. First, a narrow region of a Faraday depth spectrum with an anomalously low amplitude surrounded by broad higher amplitude features will appear, after the high-pass filter removes the broad features, as a peak with the opposite phase (perpendicular polarization angle) in the observed spectrum. The second scenario occurs when a broad feature has a sharp edge, which can occur in a Burn slab configuration \citep{Burn1966}, when the parallel magnetic field reverses sign \citep[i.e., a Faraday caustic, ][]{Bell2011}, or when the Faraday depth converges to a fixed value at large distances \citep[e.g., some of the models investigated by][]{Beck2012}. These features take the shape of a asymmetric peak, with a sharp edge on one side and a gradual tail or other broad feature on the other.

In all of these cases, a sharp change in the amplitude of the Faraday depth spectrum is required, so to understand the physical origins of such changes we must consider the various physical parameters that go into the amplitude and how they may have strong, localized variations. In the numerator of Eq.~\ref{eq:amplitude} is the polarized emissivity, which we can decompose into three terms: the total-intensity synchrotron emissivity ($\varepsilon$), the fractional polarization ($\Pi$), and a phasor defining the emitted polarization angle ($\theta_0$) as perpendicular to the magnetic field projected on the sky-plane. The synchrotron emissivity can be further decomposed, as it is proportional to the density of cosmic ray electrons ($n_{cr}$) and the strength of the magnetic field component perpendicular to the line of sight ($B_\perp$). The fractional polarization is determined by the degree of order in the magnetic field, on the scale of the angular resolution, making it affected by both physical effects and observational parameters. The polarization angle does not affect the amplitude (except through interference of overlapping Faraday depth features) so I will not consider it further. The resulting equation looks like 
\begin{equation}\label{eq:factors}
\tilde{F}(\phi) = \sum_{x_\phi} \frac{\Pi e^{2i\theta_0(x_\phi)}\varepsilon(n_{cr}(x_\phi),B_\perp(x_\phi))}{n_e(x_\phi) \left|B_\parallel(x_\phi)\right|}.
\end{equation}

Equation~\ref{eq:factors} has three important variables in the numerator: the fractional polarization (which depends on the degree of order in the magnetic field), the cosmic ray electron density, and the strength of the perpendicular component of the magnetic field, and two parameters in the denominator: the density of free (thermal) electrons and the strength of the parallel component of the magnetic field. To produce a sharp amplitude change, one or more of these parameters must change significantly in a localized region. More specifically, the region over which these parameters are significantly different must have internal Faraday rotation less than the largest measured scale in any observation measuring the polarized feature. Following \citet{VanEck2017a}, I consider a few examples of physical conditions under which these conditions could be satisfied and a Faraday-thin feature observed.

Shocks oriented perpendicular to the line of sight are a potential candidate for producing Faraday-thin features. Such shocks could significantly enhance the perpendicular component of the magnetic field, increasing synchrotron emissivity, and may also increase the degree of order of the field. Shocks are also thought to serve as sites for acceleration of cosmic rays \citep[and references therein]{Bell2013}, increasing the density of cosmic ray electrons. However, shocks also have increased matter density over their surroundings, which would increase the Faraday rotation. As a result, the effect of a shock on the Faraday depth spectrum may be complicated and requires more careful modelling to determine if a Faraday thin feature would be produced.

To produce high-amplitude, Faraday-thin Faraday depth features, conditions under which the denominator becomes very small are an attractive possibility. The strength of the parallel magnetic field is an obvious candidate, as it is very easy to imagine magnetic field configurations with regions where the magnetic field is nearly perpendicular to the line of sight. Regions where the magnetic field reverses sign along the line of sight will produce Faraday caustics \citep{Bell2011} which have a high amplitude feature with a sharp edge. Faraday caustics will generally not have an observational signature in any other tracer, but can be identified in observations with sufficiently high fractional bandwidth. \citet{Bell2011} also proposed that Faraday caustics will occur in sheets, potentially producing diffuse polarized features with a large angular size.

Regions of very low thermal electron density can also lead to high-amplitude Faraday thin features. Such regions can occur in two forms: those occupied by hot, low-density material (i.e., the hot ionized phase of the ISM, and the intergalactic medium), and those occupied by higher density, very low ionization fraction material (i.e., the warm neutral phase of the ISM). The hot, thin phase of the ISM will occupy large parts of the ISM, particularly in the Galactic halo, but these regions will often also have reduced synchrotron emissivity and large path lengths which may prevent a sharp peak from forming in some cases. Neutral regions, in the form of warm neutral clouds, may form such peaks if they have a sufficiently low ionization fraction and/or a sufficiently low density to prevent significant internal Faraday rotation while still having enough path length to accumulate a significant amount of synchrotron emission; \citet{VanEck2017a} used this to explain the the polarization features seen in their field. The cold neutral phase, specifically molecular clouds, are probably less likely to produce Faraday-thin features, on account of their very high density offsetting the low ionization fraction and leaving them with a significant free electron column, which combined with the enhanced magnetic field found in those regions results in significant internal Faraday rotation \citep{Tahani2018}.

\subsection{Other wavelength-dependent depolarization effects}\label{sec:depol}
The sensitivity to only Faraday-thin features, or equivalently the depolarization of Faraday-broad features, is the wavelength-dependent depolarization effect most often considered in the context of RM synthesis. However, there are several additional effects that may also be amplified at low-frequency that should be considered.

One is the effects of beam depolarization, where the Faraday depth profiles for all the lines of sight within the solid angle of the observing beam are averaged together and can potentially cause partial cancellation. Some portion of this can be caused by changes in the intrinsic (pre-Faraday rotation) polarization angle of the emission, which will be wavelength-independent, while the rest of the depolarization will occur by changes in the Faraday rotation across the beam, and will correspondingly depend on wavelength. This effect can be interpreted in two ways: in the Faraday depth domain, the beam is averaging the Faraday depth profiles over some solid angle, which can cause a spatially-varying Faraday-thin feature to become sufficiently Faraday-thick on the scale of the beam as to depolarize significantly; alternatively, in the wavelength domain a spatially-varying Faraday depth will cause the polarization angle at a given frequency to vary significantly across the beam, with longer wavelengths having stronger variations in polarization angle due to experiencing more Faraday rotation.

Spatial variations in the Faraday depth of a polarized feature have the net effect of driving polarized emission from large angular scales to smaller scales \citep{Schnitzeler2009}. This has two main effects: the first is the previously defined beam depolarization, where polarized emission is moved to small angular scales which are not measured (in the case of interferometers, due to the lack of sufficiently long baselines); the second is the reduction in polarized emission on the largest angular scales. As a result, at very low frequencies (i.e., the LOFAR and MWA regime) the effect of missing short-spacing data may not be a major problem for large-area Faraday tomography with interferometric observations, but this effect has not yet been quantified.

Wavelength dependent depolarization can also be strongly affected by the specific shape of the Faraday depth profile. The simplest example of this is comparing a Burn slab \citep{Burn1966}, which has a top-hat Faraday depth profile and a corresponding $\sigma_\phi\lambda^{-2}$ depolarization law (where $\sigma_\phi$ is the characteristic width of the profile), while a Gaussian Faraday depth profile, which can be caused by a turbulent Faraday rotating screen \citep{Burn1966}, has  a $e^{-\sigma_\phi^2\lambda^4}$ depolarization law which falls off much more rapidly with increasing wavelength. In the Faraday depth domain, this can be interpreted in terms of the `high-pass filter' effect of RM synthesis, as a Gaussian is dominated by the broad Fourier components while the top-hat function has significant emission in narrow components to produce the sharp edges. As a result, different distributions with the same $\sigma_\phi$ may experience very different depolarization.

\section{Recent Observations}\label{sec:current}
The earliest low-frequency Faraday tomography was performed using observations from the WRST at frequencies in the 115-175 MHz and 310-383 MHz ranges \citep[e.g.,][]{deBruyn2005,Schnitzeler2009,Brentjens2011,Pizzo2011}. These observations were primarily targeted on galaxy clusters, but also identified diffuse polarization features which they attributed to the Milky Way. \citet{deBruyn2005} and \citet{Brentjens2011} used 300-MHz WRST data to study the Perseus cluster, and identified several polarized features on large (10\arcmin) angular scales, which were initially attributed to the cluster \citep{deBruyn2005} but were later though to be within our Galaxy \citep{Brentjens2011}. \citet{Pizzo2011} used WRST data in several frequency ranges from 115-1795 MHz to observe the Abell 2255 cluster, and despite deliberately filtering out large baselines to remove Galactic contributions they identified polarized features that they attributed to our Galaxy. \citet{Schnitzeler2009} used WSRT data in the 350 MHz band to observe Galactic emission in a large field. \citet{Iacobelli2013} also used low-frequency (139--155 MHz) WSRT data to observe the FAN region, an area of sky previously known to have unusually high polarized intensity; they identified several distinct polarized features with small separations in Faraday depth, and attempted to associate them with known structures in the ISM. Many of these studies had limited fields of view, opening questions about the full extent the observed structures and how much of the sky is occupied by such features, and also had limited short-baseline coverage, opening question about whether features on larger angular scales were being missed.

With the construction of LOFAR and MWA, a new capability was opened up to explore polarized emission on larger angular scales and larger regions of the sky. \citet{Jelic2014}, \citet{Jelic2015}, and \citet{VanEck2017a} applied Faraday tomography to individual LOFAR observations and found many diffuse polarized features on large scales. The 3C196 field of \citet{Jelic2015} yielded a number of interesting features: bright diffuse Galactic polarization was observed, but with one long narrow filament clearly offset in Faraday depth from its surroundings, indicating a filamentary foreground screen; an alignment between low-frequency polarized features and dust polarization \citep{Zaroubi2015}; and a series of long and very straight depolarization canals \citep{Jelic2018}. \citet{VanEck2017a} found two distinct polarized features in their field, overlapping on the sky but separated in Faraday depth by less than 10 \radu, and and associated these polarized features with neutral clouds within 1 kpc of the Sun.

\citet{Lenc2016} applied Faraday tomography to MWA observations over a very large region (625 deg$^2$) near the South Galactic Pole at low resolution (50\arcmin). They found diffuse polarized emission in their field at a high intensity level (4--11 K polarized brightness temperature at 154 MHz), indicating that this region had either strong polarized emission or that beam depolarization was not extreme in that region. They found that this emission was also very local, within 220 pc, although this assumed a uniform Faraday rotating medium and did not account for the presence of the Local Bubble, which is expected to have very low Faraday rotation.

\begin{figure}[htbp]
   \centering
   \includegraphics[width=\textwidth,height=0.9\textheight,keepaspectratio]{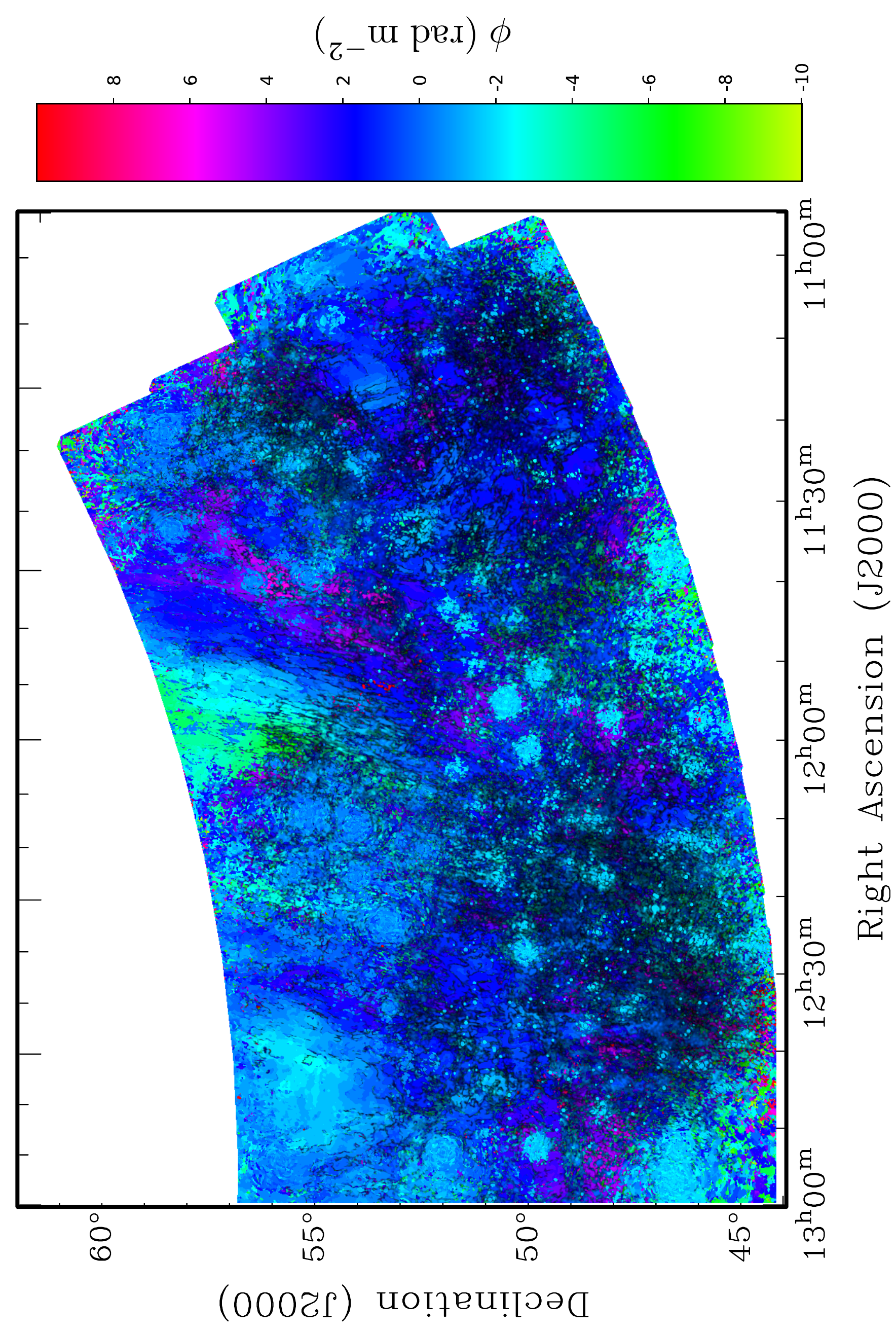} 
   \caption{A map of Faraday depth of the strongest peak (encoded as hue) and peak polarized intensity (encoded as brightness, capped at 3 mJy PSF\inv\ RMSF\inv), calculated from the western region of the Faraday depth cube of \citet{VanEck-thesis}. Continuous gradients in Faraday depth can be observed in several regions. Instrumental leakage of bright point sources shows up as small bright-blue features throughout the image.}
   \label{fig:mosaic_right}
\end{figure}

\begin{figure}[htbp]
   \centering
   \includegraphics[width=\textwidth,height=0.9\textheight,keepaspectratio]{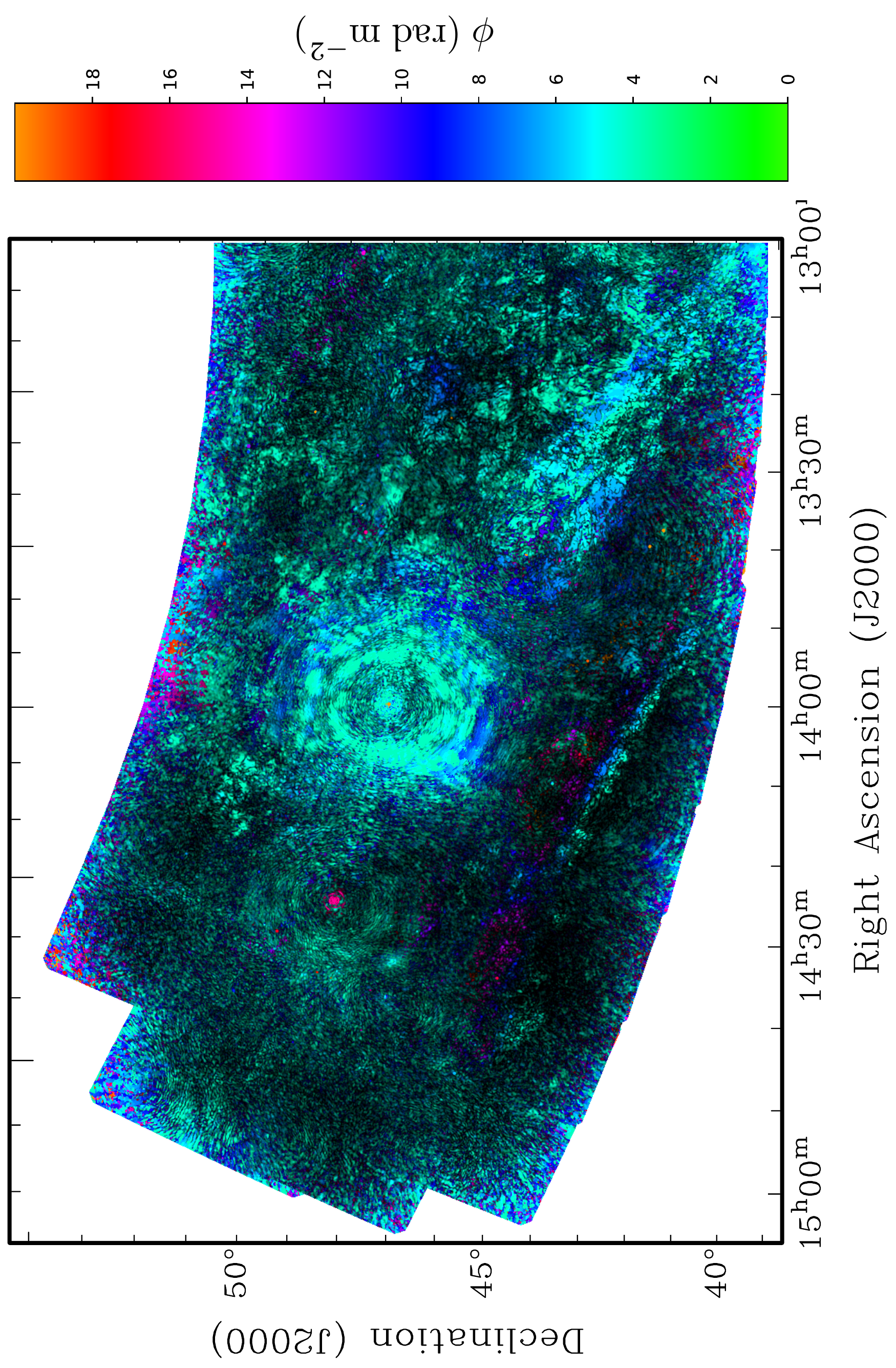} 
   \caption{As Figure~\ref{fig:mosaic_right}, but for the eastern region of the mosaic. The instrumental leakage from 3C295 dominates the center, but patches of diffuse polarized emission can be seen on the right and a long, straight feature can be seen below the center, which traces the outer edge of a neutral hydrogen filament (not shown).}
   \label{fig:mosaic_left}
\end{figure}

\citet{VanEck-thesis} combined Faraday depth cubes from 60 LOFAR observations from LOTSS to create a 568 deg$^2$ mosaic covering a region in the northern Galactic halo. A collapsed version of this cube, showing the Faraday depth and polarized intensity of the brightest peak in the Faraday depth spectrum of each pixel, is shown in Figures~\ref{fig:mosaic_right} and \ref{fig:mosaic_left}. In this mosaic there are several polarized features that extend beyond ten degrees in length, much larger than the field of view of a single LOFAR observation. Many of these features could be clearly seen as continuous sheets of (Faraday-thin) polarized emission, covering tens of square degrees, with gradients in the Faraday depth of the peak. \citet{VanEck-thesis} proposed that the origin of the Faraday-thin emission may be Faraday caustics, and that the gradients in Faraday depth (especially where the sign of Faraday depth changes) could be caused by loops in the magnetic field, but they also acknowledged that other interpretations are possible.

For Faraday tomography of extragalactic objects, such as nearby galaxies, low-frequency data has not been as successful. \citet{Mulcahy2014} and \citet{Farnes2013} used observations from LOFAR at 150 MHz and the Giant Meterwave Radio Telescope at 610 MHz respectively to search for polarized emission from M51, but did not detect significant polarization. Faraday tomography has been successfully applied to nearby galaxies at higher frequencies \citep[e.g.][]{Mao2012a}, so this is a sign of strong wavelength dependent depolarization. Based on the results from Milky Way Faraday tomography, it seems quite reasonable to think that Faraday thin features exist in other galaxies, but gradients in Faraday depth will cause them to appear as broad features when averaged over the beam, causing them to be appear depolarized at low frequencies.

\section{Future Observations}\label{sec:future}
\begin{figure}[htbp]
   \centering
   \includegraphics[width=\textwidth,height=0.9\textheight,keepaspectratio]{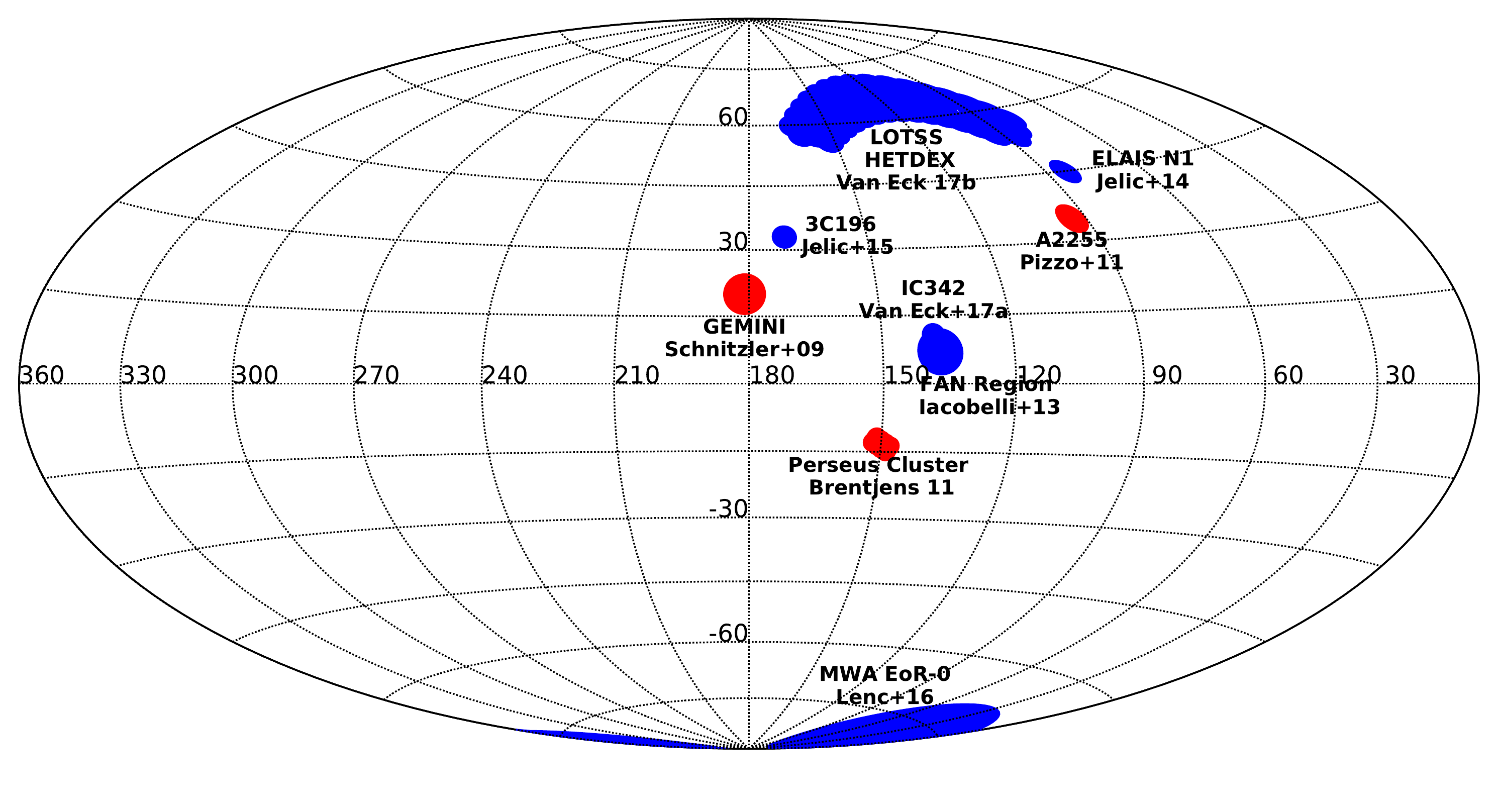} 
   \caption{A map of the sky, in a Hammer projection of Galactic coordinates, showing the regions covered by current low-frequency Faraday tomography. Observations below 250 MHz are shown in blue, and observations between 250 and 1000 MHz are shown in red (this considers only the lowest frequency included in the RM synthesis, so some fields will have included data above 1000 MHz). Combined, these observations cover less than 4\% of the total sky.}
   \label{fig:skymap}
\end{figure}

The observations to date have shown low-frequency Faraday tomography can be highly successful in finding polarization features in the ISM of the Milky Way. However, the parameter space of sky area, Faraday depth resolution, and sensitivity to Faraday-broad features is still very sparsely explored. Figure~\ref{fig:skymap} shows the regions of the sky that have been mapped to date, which represents a small fraction of the total sky.

Both LOFAR and MWA are conducting all-sky total intensity surveys (LOTSS and GLEAM/GLEAM-X, respectively), and the data for each has been shown to be suitable for polarization analysis \citep[and Riseley et al, submitted, respectively]{VanEck-thesis}. As a result, there is the opportunity to conduct commensal polarization and Faraday tomography surveys with these data. Together these would give full sky coverage at 150 MHz with arcminute angular resolution and baselines down to a few tens of meters or less. 

Another avenue that is being explored is single-dish polarization surveys, which are sensitive to large-scale Galactic features. The Global Magneto-Ionic Medium Survey \citep[GMIMS, ][]{Wolleben2009} is an ongoing project to observe the entire sky from approximately 300 to 1800 MHz, through 6 sub-surveys dividing the frequency range into three segments and the sky into the northern and southern hemispheres. The first two sub-surveys, covering the northern hemisphere from $\sim$1.3--1.8 GHz and the southern hemisphere from 300--480 MHz, are expected to be released soon (T.~Landecker, private communication). While GMIMS will suffer from limited angular resolution due to the limits of single-dish and compact interferometer telescopes at low frequencies, the very wide frequency coverage will offer unprecedented opportunity to explore Faraday complexity, by giving both the Faraday depth resolution of the lowest frequencies and the sensitivity to broad features of the higher frequencies.

Low-frequency Faraday tomography of nearby galaxies will remain challenging, as very high angular resolution will be required to overcome beam depolarization of Faraday thin features. The LOTSS data may be the most promising possibility in the near future, as it can achieve 5\arcsec\ resolution, but further progress may have to wait for the Square Kilometre Array to provide higher resolution and sensitivity at low frequencies.

\section{Summary}\label{sec:summary}
Faraday tomography is a very powerful tool for studying cosmic magnetic fields because it provides us with information on the distribution of Faraday rotation along a line of sight. Using Faraday tomography with low-frequency radio polarization data enhances this advantage, by providing improved Faraday depth resolution, but this comes with the cost of not detecting features that are broadened in Faraday depth. However, this limitation can be exploited as it provides constraints on the thermal electron and magnetic field distributions that can lead to a detectable low-frequency polarization feature. In particular, low-frequency Faraday tomography is sensitive to polarized emission from comparatively small regions (required to produce Faraday thin features), making it particularly effective in probing structure in the magnetic field on smaller spatial scales when compared with higher frequency data where the entire line of sight is blended together.

Recent results have demonstrated our capability to detect and analyze low-frequency polarization with new instruments such as LOFAR and MWA. These observations have found a number of interesting, previously known polarization features of predominantly Galactic origin, and have shown that some of these features have very interesting relationships with other aspects of the ISM.

However, we have only scratched the surface of exploring the low-frequency Faraday depth sky, as most of the sky remains unexplored. Future polarization surveys with LOFAR and MWA have the potential to cover the whole sky at 150 MHz and arcminute to arcsecond resolution, while GMIMS will cover the whole sky from 300--1800 MHz at degree resolution. These surveys, once complete, will dramatically advance our knowledge of polarized emission, Faraday rotation, and the associated magnetic fields in our Galaxy and in extragalactic radio sources.

\section*{Acknowledgements}
\small
The author wishes to thank Marijke Haverkorn and the three reviewers for their comments on the manuscript.\\

The preparation of this manuscript made use of Astropy, a community-developed core Python package for Astronomy \citep{Astropy}; NumPy \citep{Numpy}; IPython \citep{Ipython}; and matplotlib \citep{Matplotlib}.

\bibliographystyle{aa}
\bibliography{References}

\end{document}